\documentclass[sigconf]{acmart}

\copyrightyear{2024}
\acmYear{2024}
\setcopyright{rightsretained}
\acmConference[SPAA '24]{Proceedings of the 36th ACM Symposium on Parallelism in Algorithms and Architectures}{June 17--21, 2024}{Nantes, France}
\acmBooktitle{Proceedings of the 36th ACM Symposium on Parallelism in Algorithms and Architectures (SPAA '24), June 17--21, 2024, Nantes, France}\acmDOI{10.1145/3626183.3659977}
\acmISBN{979-8-4007-0416-1/24/06}
\acmDOI{10.1145/3626183.3659977}

\usepackage{epigraph}
\usepackage{tlatex}
\usepackage{csquotes}
\usepackage[ruled,vlined,linesnumbered]{algorithm2e}
\usepackage{multicol,multirow,tabularx}
\usepackage{listings}
\usepackage{tikz}
\usepackage{soul}
\usepackage{listings}
\usepackage{subcaption}
\usepackage{algpseudocode}
\usepackage{todonotes}

\usetikzlibrary{positioning,decorations.pathreplacing}
\newcommand{\code}[1]{\texttt{#1}}

\AtBeginDocument{%
  \providecommand\BibTeX{{%
    \normalfont B\kern-0.5em{\scshape i\kern-0.25em b}\kern-0.8em\TeX}}}

\begin{document}

\title{ALock: Asymmetric Lock Primitive for RDMA Systems}
\subtitle{Technical Report}

\author{Amanda Baran}
\email{adb321@lehigh.edu}
\affiliation{%
  \institution{Lehigh University}
  \city{Bethlehem}
  \state{PA}
  \country{USA}
}

\author{Jacob Nelson-Slivon}
\email{jjn217@lehigh.edu}
\orcid{1234-5678-9012}
\affiliation{%
  \institution{Lehigh University}
  \city{Bethlehem}
  \state{PA}
  \country{USA}
}

\author{Lewis Tseng}
\email{LTseng@clarku.edu}
\affiliation{%
  \institution{Clark University}
  \city{Worcester}
  \country{MA}
  \country{USA}
}

\author{Roberto Palmieri}
\email{palmieri@lehigh.edu}
\affiliation{%
  \institution{Lehigh University}
  \city{Bethlehem}
  \state{PA}
  \country{USA}
}

\renewcommand{\shortauthors}{Baran et al.}

\begin{abstract}
Remote direct memory access (RDMA) networks are being rapidly adopted into industry for their high speed, low latency, and reduced CPU overheads compared to traditional kernel-based TCP/IP networks. 
RDMA enables threads to access remote memory without interacting with another process. 
However, atomicity between local accesses and remote accesses is not guaranteed by the technology, hence complicating synchronization significantly. 
The current solution is to require threads wanting to access local memory in an RDMA-accessible region to pass through the RDMA card using a mechanism known as loopback, but this can quickly degrade performance. 
In this paper, we introduce ALock, a novel locking primitive designed for RDMA-based systems. ALock allows programmers to synchronize local and remote accesses \textit{without} using loopback or remote procedure calls (RPCs). 
We draw inspiration from the classic Peterson's algorithm to create a hierarchical design that includes embedded MCS locks for two cohorts, remote and local. 
To evaluate the ALock we implement a distributed lock table, measuring throughput and latency in various cluster configurations and workloads. 
In workloads with a majority of local operations, the ALock outperforms competitors up to 29x and achieves a latency up to 20x faster.
\end{abstract}


\maketitle

\section{Introduction}
Coordinating access to shared resources is a fundamental challenge in concurrent computation and has motivated the widespread adoption of hardware enabled synchronization mechanisms like compare-and-swap (CAS), whose importance in concurrent computing is indubitable. 
An atomic CAS operation enables an arbitrary number of threads to agree on a value in a wait-free manner~\cite{herlihy91waitfree}, making it a powerful tool for building mutual exclusion primitives.
The value of atomic CAS operations is likewise evident in the increasingly popular network communication technology, remote direct memory access (RDMA).

Remote direct memory access is a popular network communication technology that aims at implementing the shared-memory abstraction in the distributed setting by allowing a thread to access memory on a remote machine \textit{without} interacting with another process~\cite{technologies2015rdma, 2007infiniband, gzc2016gen}.
In addition to its ability to read and write memory on another machine, RDMA also enables threads to perform atomic read-modify-write (RMW) operations on remote memory, like compare-and-swap.

An important factor to consider when dealing with RDMA operations is that the atomicity between local (i.e., shared-memory) and remote (i.e., RDMA) accesses \textit{is not guaranteed}. That means:
\begin{itemize}
    \item[\textit{i)}] RDMA reads and writes are only atomic with their local counterparts for accesses within a single cache line~\cite{dragojevic2014farm} (typically only 64 bytes); and
    \item[\textit{ii)}] RDMA RMW operations are \textit{not} atomic with local RMW operations.
\end{itemize}
Essentially, from the perspective of local memory, a remote RMW is nothing more than a read followed by a write to the same memory location.
As a result, the lack of atomicity for RMW operations makes it significantly more difficult to synchronize local and remote accesses.

In practice, RDMA RMW operations are frequently used in RDMA-based distributed systems when synchronizing concurrent accesses (e.g., ~\cite{chen2017fast, binnig2016end, zamanian2017end, yoon2018distributed}).
To ensure atomicity between operations in these systems,
threads performing local accesses must use the \emph{loopback} mechanism, which allows a thread to access RDMA memory on its own machine by passing through the local RDMA network interface controller (RNIC)~\cite{zehnder2015scalable}.
Although fast compared to other network communication, RDMA is still at least an order of magnitude slower than shared memory operations~\cite{kalia2016design, zehnder2015scalable}, and suffers from non-uniform memory access (NUMA) behaviors~\cite{nelson2020performance}, degrading performance for local accesses.
Additionally, the RDMA loopback mechanism can cause poor performance due to internal congestion~\cite{kong2022collie}. In Section~\ref{sec:issue}, we discuss an empirical study we conducted to confirm our claims about loopback congestion and the scalability issues of RDMA.

Alternatively, systems could leverage remote procedure calls (RPCs) to allow all synchronization to be handled exclusively by threads on receiving nodes.
However, RPCs follow the traditional send/receive programming model, which has been shown to nullify the performance benefit of directly accessing remote memory~\cite{taleb2018tailwind}.
However, in practice, RPCs are still used in RDMA-based systems (e.g.,~\cite{dragojevic2014farm, ziegler2019designing, kalia2016fasst, mitchell2013using, wang2015hydradb}), and their prevalence is attributed in part to the challenges associated with synchronizing local and remote accesses.

In this paper, we describe a new locking primitive, \textit{ALock}, designed to capture the nuanced requirements of synchronizing local and remote accesses in RDMA-aware systems \underline{\textit{without}} using loopback or RPCs.
In our solution, we enforce that threads performing local accesses only use local (i.e., shared-memory) operations. However, since atomicity is not guaranteed among local and remote (i.e., RDMA) RMW operations, we require a new mutual exclusion algorithm between these two groups of operations.

\textbf{Our Design}. The intuition behind the ALock design begins with reducing the problem to a two-process mutual exclusion. 
A natural choice is to find inspiration from Peterson's algorithm~\cite{peterson1981myths}, a well-known, starvation-free, and fair mutual exclusion protocol for two processes using only shared memory to communicate (more details about Peterson's algorithm can be found in Section~\ref{sec:ogpete}).


To allow for \textit{multi}-thread synchronization, we extend Peterson's algorithm to include two \textit{cohorts} (as opposed to just two threads), \textit{remote} and \textit{local}.
Threads within each cohort compete amongst each other to determine a \textit{leader}. 
The leaders of the two cohorts then compete using our modified version of Peterson's lock algorithm to successfully acquire the ALock, finally entering the critical section.
Thanks to this hierarchical design, threads within each cohort can use APIs that are guaranteed to be atomic with each other. 
Specifically, threads performing local accesses can use shared-memory operations, and threads performing remote accesses can use RDMA operations. Notably, there is no need to use the loopback mechanism anymore.

In addition to eliminating loopback for local accesses, it is critical to limit the number of RDMA operations for remote accesses in order to prevent congestion in the RNIC, which causes performance deterioration (see Section~\ref{sec:issue}). For this reason, we take advantage of the design of the widely used MCS queue lock~\cite{mellorcrummey1991algorithms}, which allows threads to spin locally while waiting in a queue to acquire the lock.
This combination of locks has similar characteristics to lock cohorting~\cite{dice2012lock}, which decouples the synchronization within a cohort with the synchronization among cohorts.

To evaluate the ALock, we build a distributed lock table and measure the throughput and latency on multiple cluster configurations under various contentions, ratios between local and remote accesses, and workloads.
We compare against the commonly used RDMA CAS-based spinlock and an RDMA-enabled MCS Queue lock. 
In the presence of a majority of local operations, ALock outperforms competitors up to 29x. 
As a result, data repositories that use one-sided RDMA operations and currently rely on specialized hardware or loopback to achieve atomicity between remote and local operations can relax these constraints and improve performance with ALock.

To the best of our knowledge, we are the first to fully develop a solution that solves mutual exclusion specifically for RDMA in a manner that does \textit{not} require involving the loopback RNIC or RPC handling for local accesses, is \emph{starvation-free}
and \emph{fair}.

The ALock is released as an open-source project and can be found at~\cite{alockcode}. A preliminary design of this work appears in ~\cite{nelson2022discba}.

The formal TLA+ specification of the ALock's correctness, liveness, and fairness can be found in Appendix~\ref{app:tla}.

\section{RDMA Scalability and pitfalls}
\label{sec:issue}

RDMA is a technology increasingly affecting the design of many distributed systems due to its unquestionably fast operations. 
These operations include APIs to send/receive messages but also to operate directly on remote memory (named one-sided operations). 
The presence of these one-sided operations is considered RDMA's true breakthrough because it challenges the message-passing model, which is typically used to design distributed interactions. 
Although promising as a technology, RDMA's programmability still requires innovations, such as our ALock, as its performance scalability is hampered by two factors, namely the loopback network congestion and QP thrashing~\cite{wang2021star, kong2022collie}.

\begin{figure}[ht] 
    \centering
    \includegraphics[scale=.5]{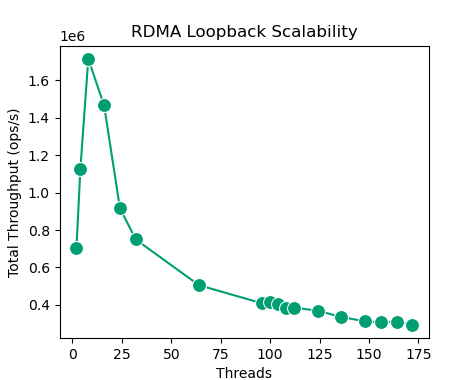}
    \caption{RDMA Spinlock with 1k locks on 1 Node.}
    \label{fig:spin}
\end{figure}

To assess the performance of RDMA loopback and show the network saturation pitfalls it comes with, we perform a simple experiment to verify our intuition.
We run a simple spinlock algorithm using 1000 locks, which shows low to no logical contention, on a single machine equipped with an Intel Zeon E5-2450 processor with 8/16 cores/threads and one Mellanox dual port CX3 RNIC. 
As shown in Figure~\ref{fig:spin}, the peak throughput is reached at a few threads.
After that, the loopback traffic drains the PCIe bandwidth, causing accumulation in the RNIC's RX buffer and slowing down the CAS operations despite the lack of logical contention.

Another well-documented problem with RDMA in large-scale data centers is connection scalability. 
In order for one-sided RDMA operations to avoid the involvement of the host CPU, the RNIC also manages the connection context. 
This connection context, sometimes referred to as QP Context (QPC), maintains all the attributes required for both sending data and keeping track of the connection state. 
Typical commercial RNICs such as the ConnectX-4 require 256 bytes to maintain the QPC for each RDMA connection \cite{mellanoxPRM}.
The Infiniband specification states that the RNIC can handle at most $2^{24}$ QPCs~\cite{2007infiniband}.
However, at 256 bytes each, this would require the RNIC to have 4GB of space to store QPCs for all $2^{24}$ connections.
In reality, the RNIC has a very small cache on its chip, which is not nearly enough to maintain all of its connections. 
Recent work has shown that the message rate of commodity RNICs declines after 450 connections \cite{wang2021star}. 
As a result, RDMA-enabled systems can experience a side effect known as \textit{QP thrashing}. 
This occurs when the number of QPs being used does not fit in the RNIC cache, so the RNIC is continuously loading and evicting QPCs from the cache, quickly degrading performance. 
Systems using RDMA loopback for local accesses can still be affected by QP thrashing regardless of whether the memory is local to the requesting thread. 

The design of the ALock directly solves the first issue of using loopback in the presence of workloads accessing local memory. Also, it limits QP thrashing by removing $\frac{1}{n}$ QPs from the system (where $n$ represents the number of nodes) if we assume an identical workload on each node.


\section{Background}

\subsection{Peterson's Algorithm} \label{sec:ogpete}
Peterson's algorithm (also referred to as Peterson's lock) by Gary L. Peterson~\cite{peterson1981myths} is a well-known algorithm for mutual exclusion that is both starvation-free and fair. 
It only relies on read and write operations to coherent shared memory to synchronize.
Since atomic read and write operations can be used over remotely accessible memory, Peterson's algorithm can also be implemented directly over RDMA, with the appropriate memory fences, to coordinate access between a local and remote thread.

Peterson's algorithm uses two variables: a boolean array \code{flag} of size 2 and an integer \code{victim}. 
A process announces its interest in entering the critical section by setting its flag to \code{true} and setting itself as the victim. 
By doing so, the algorithm guarantees fairness as the waiting process will always access the critical section next, regardless of any performance differences between the two threads. 
After completing the critical section, the process sets its flag to \code{false} to indicate it does not want to execute anymore. 
The waiting thread waits until the executing process sets its flag to \code{false}, or the thread is no longer the victim. 
By alternating access to the critical section through the use of the "victim" variable, the algorithm ensures both processes eventually gain access.

\subsection{MCS Lock}
Mellor-Crummy and Scott presented a scalable mutual exclusion algorithm for multiprocessors in 1991, the MCS lock~\cite{mellorcrummey1991algorithms}. 
The algorithm is fair because it is equivalent to a FIFO-pattern queue. 
It is also contention-free, as each processor spins on a local variable.
This property is desirable when extending the MCS lock to be used in a distributed system, as remote spinning severely limits performance due to network congestion.

A process interested in acquiring the lock first allocates a locally-accessible descriptor, which contains a boolean flag to spin on and a pointer to form the queue.
To acquire the lock, the process adds its descriptor to the end of the queue. 
If the process has a predecessor, the process spins on its local boolean variable until its predecessor passes the lock by writing to the variable. 
Otherwise, the process is the head of the queue and is the lock holder.
To release the lock, the process writes to the boolean variable of its successor.
If the process has no successor, the tail is set to \code{NULL}.

\section{System Model} \label{sec:model}

We model an RDMA-based distributed system as a set of nodes $N$ and threads\footnote{The theory literature often refers to threads as processes. 
These names are interchangeable in our context.} $T$, for which $t_i^j$ is a thread running on node $n_i$ with identifier $j$.
All threads can access an RDMA-accessible shared-memory partitioned among the nodes.
In our model, all data and metadata are stored in RDMA-accessible memory.
Similar to previous RDMA system models~\cite{aguilera2018passing}, we assume that threads are asynchronous and that accesses to memory are failure-free.

The notion of locality is crucial in our model. 
Although not new~\cite{hendler2013lilac}, we define it in relation to the APIs a thread uses to operate on RDMA-accessible memory.
Definitions~\ref{def:local} and~\ref{def:remote} define local and remote access.

\begin{definition}[Local Access] \label{def:local}
    A thread $t_i^j$, executing on node $n_i$, performs a \textit{local access} using shared-memory operations if the RDMA-accessible memory being accessed is stored on $n_i$ . 
\end{definition}

\begin{definition}[Remote Access] \label{def:remote}
    A thread $t_i^j$, executing on node $n_i$, performs a \textit{remote access} using RDMA operations if the RDMA-accessible memory being accessed is \textit{not} stored on $n_i$ . 
\end{definition}

For each class of access (local or remote), memory is accessed through three operations: read, write, and compare-and-swap.
We denote the shared-memory (or local) operations using \code{Read}, \code{Write} and \code{CAS}, and the RDMA (or remote) operations with \code{rRead}, \code{rWrite} and \code{rCAS}.
As shown in Table~\ref{tab:atomicity}, the atomicity of operations between classes is \textit{not} guaranteed due to the design of RDMA.

\begin{table}[ht]
    \centering
    \begin{tabular}{|cc|ccc|}
        \hline
        \multicolumn{2}{|c|}{\multirow{2}{*}{Access (8B)}} & \multicolumn{3}{c|}{Remote (RDMA)} \\
        \cline{3-5}
        & & \code{Read} & \code{Write} & \code{CAS}  \\
        \hline
        \multirow{3}{*}{\rotatebox[origin=c]{90}{Local}} & \multicolumn{1}{|c|}{\code{Read}} & Yes & Yes & Yes \\
        & \multicolumn{1}{|c|}{\code{Write}} & Yes & Yes & {No} \\
        & \multicolumn{1}{|c|}{\code{RMW}} & Yes & Yes & {No} \\
        \hline
    \end{tabular}
    \caption{Atomicity between 8-byte local and remote accesses.}
    \label{tab:atomicity}
\end{table}

Since remote operations complete asynchronously, local and remote access to a given memory location may be reordered~\cite{dan2016modeling}.
As a result, an \code{rCAS} operation appears to a thread performing local accesses as if it were a \code{Read} followed by a \code{Write}. Hence, interleaves with other local operations are possible.
Thus, the programming model requires that programmers wait until operations are complete by using appropriate memory fences to guarantee correct ordering~\cite{2007infiniband}.

We say a memory object is \emph{operation-asymmetric} if, given two threads, the intersection of the operations they can perform on that object is \emph{not equal} to their union.
ALocks are designed to deal with the operation asymmetry of RDMA-accessible memory, hiding the complexity from the programmer while simultaneously optimizing for it.

\section{Alock Algorithm and Design}\label{sec:alock}

\SetKwProg{MyStruct}{Struct}{ contains}{end}

ALock innovates over the idea of lock cohorting~\cite{dice2012lock} by defining cohorts with respect to the APIs used by threads operating on the lock. 
More specifically, in our case, two cohorts are defined {as follows: one cluster of threads performing local accesses, and one cluster of threads performing remote accesses.

The ALock is a composition of two of our modified MCS queue locks, one for each cohort, and
our modified Peterson's algorithm. 
Elements in the MCS queues of the above MCS locks are made of descriptors, which are memory regions describing the state of the lock request. 
Since lock requests can be generated by threads from any node in the system, these MCS queues are effectively linked to possibly distributed memory. For simplicity, we refer to each MCS queue by its \textit{tail} (i.e., the pointer to the last element of the queue).

To maximize performance and reduce the amount of metadata required, the remote and local tails ($tail\_r$ and $tail\_l$, respectively) are also interpreted as the status flags used in Peterson's algorithm~\cite{herlihy2021}. 
Recall from Section~\ref{sec:ogpete} that, when the flag is set to \code{true} for process $i$, that means $i$ is either interested in capturing the lock or is currently in the critical section. 
Similarly, in our algorithm, tails contain either a \code{NULL} pointer or a pointer to a descriptor.
The latter indicates that the remote ($tail\_r$) or local ($tail\_l$) cohort is interested in or has acquired the lock.
Embedding Peterson's flag semantics into the MCS queues avoids an additional, possibly remote, memory operation.
Also, as required for the Peterson' algorithm, we include the boolean \code{victim} field, which is used to yield the lock to a waiting thread of the other cohort. 

\textbf{Metadata.}
The metadata used by a thread to interact with an ALock is represented in Algorithm~\ref{alg:structs}. 
Two descriptors are allocated in RDMA-accessible memory, one \code{LocalDescriptor}, which is used by threads that are local to the ALock, and one \code{RemoteDescriptor}, which is used by threads that are remote to the ALock.
Each of these descriptors has a budget field associated with it. 
In short, the budget is used to introduce fairness between ALock's cohorts. 
A comprehensive discussion about that can be found later in the section.

\begin{algorithm}
    \caption{ALock Metadata}
    \label{alg:structs}
    \MyStruct{ALock}{
        rdma\_ptr<$RemoteDescriptor$> $tail\_r$\;
        rdma\_ptr<$LocalDescriptor$> $tail\_l$\; 
        int victim\;
    }
    \MyStruct{LocalDescriptor}{
        int local\_budget\;
        $LocalDescriptor*$ next\;
    }
    \MyStruct{RemoteDescriptor}{
        int remote\_budget\;
        rdma\_ptr<$RemoteDescriptor$> next\;
    }
\end{algorithm}

The ALock provides two operations, \code{Lock(rdma\_ptr<ALock>)} and \code{Unlock(rdma\_ptr<ALock>)}, shown in Algorithm~\ref{alg:lock}. 

\begin{algorithm}[h]
    \DontPrintSemicolon
    \caption{ALock Operations}
    \label{alg:lock}
    \code{Lock(rdma\_ptr<ALock>)} \label{line:lock}
    \Begin{
        \code{$passed$} $\gets$ qLock(rdma\_ptr<ALock>) \\ \label{line:qlock}
        \If{passed $=$ false}{ 
            \code{pReacquire(rdma\_ptr<ALock>)}        
        }
    }
    \code{Unlock(rdma\_ptr<ALock>)}
    \Begin{
        \Return \code{qUnLock(rdma\_ptr<ALock>)} 
    }
\end{algorithm}

\textbf{Lock Procedure}.
Upon initiating a \code{Lock(rdma\_ptr<ALock>)} operation, it is first determined whether the thread is performing a local (see Definition ~\ref{def:local}) or remote (see Definitions~\ref{def:remote}) access on the requested lock. 
This is done by checking the ID of the node where the lock is located, which is embedded in the first 4 bits of the RDMA pointer.
Based on the classification of the lock access request, the thread first competes to be the leader of either the remote or local cohort for that lock via our modified remote or local MCS Queue algorithm (see Algorithm~\ref{algo:rmcs}). 
In our algorithm, if the call to \code{qLock(rdma\_ptr<ALock>)} returns \code{false} on Line~\ref{line:notpassed}, that means there is no current lock holder.
Therefore, the thread needs to then compete in our modified Peterson's algorithm (see Algorithm~\ref{algo:cohorted}) in order to finally acquire the ALock.
Otherwise, if \code{qLock(rdma\_ptr<ALock>)} returns \code{true} on Line~\ref{line:passed} of Algorithm~\ref{algo:rmcs}, we say the MCS lock was \textit{passed}. 
This means that the thread did not swap onto an empty MCS queue. Instead, it waited for a predecessor to hand over the lock by spinning on a local variable in its descriptor. This local variable is written by a predecessor in order to notify the thread that it now owns the lock.

Algorithm~\ref{algo:rmcs} shows the modified MCS algorithm for remote accesses, whereby the algorithm for local accesses simply requires replacing each remote access with a local (i.e., shared-memory) one. 
For simplicity, let us assume this is the first time the lock is ever acquired. 
Because of that, \code{qLock(rdma\_ptr<ALock>)} returns \code{false}, meaning the lock was not passed, and therefore, the thread needs to participate in our modified Peterson's algorithm (Algorithm \ref{algo:cohorted}) to finally obtain the ALock.

\begin{algorithm}[ht]
    \caption{Modified Remote MCS Queue Lock}
    \label{algo:rmcs}
    \KwData{(constants) \code{kInitBudget}; (global) \code{$tail\_r$}; (process-local) \code{desc}}
    \code{qLock(rdma\_ptr<ALock>)}
    \Begin{
        \code{desc}  $\gets$ \code{RemoteDescriptor($null$)} \\
        prev $\gets$ rCAS($tail\_r$, $nullptr$, \&desc) \label{line:swap} \\
        \If{\code{prev} $=nullptr$}{
            \code{desc.remote\_budget} $\gets$ \code{kInitBudget} \\
            \Return $false$ \tcp*{Lock was not passed} \label{line:notpassed} 
        } 
        \code{desc.remote\_budget} $\gets -1$ \\
        \code{rWrite(\&(prev->next), \&desc)} \\
        \tcp{Spins locally}
        \lWhile{\upshape \code{desc.remote\_budget} $=-1$}{\textbf{wait} \label{line:wait}}
        \If{\upshape \code{desc.remote\_budget} $=0$}{\code{pReacquire()} \tcp*{Provides fairness}
        \code{desc.remote\_budget} $\gets$ \code{kInitBudget}}
        
        \Return $true$  \tcp*{Lock was passed} \label{line:passed}
    }
    \code{qUnlock(rdma\_ptr<ALock>)}
    \Begin{
        \code{curr} $=$ \code{rCAS($tail\_r$, \&desc, $nullptr$)} \\
        \If{\upshape \code{curr} $\neq$ \code{\&desc} }{
            \tcp{Wait for successor to enqueue}
            \lWhile{\upshape \code{desc.next} $=nullptr$}{\textbf{wait}} 
            \tcp{Pass the lock}  
            \code{rWrite(\&(desc.next->remote\_budget), desc.remote\_budget - 1)}
        }
    } 
    \code{qIsLocked()}
    \Begin{
        \code{\Return \code{rRead($tail\_r$)} $\ne nullptr$}
    }
\end{algorithm}

\begin{algorithm}
\caption{Modified Peterson's Lock}
\label{algo:cohorted}
\KwData{(global) \code{cohort[2]}, \code{victim}}
\code{pReacquire()}
\Begin{
    \code{id} $\gets$ \code{getCid()} \tcp*{Get ID of process cohort}
    \code{other} $\gets 1 -$ \code{id} \\
    \code{victim} $\gets$ \code{id} \tcp*{Yield lock to waiting cohort} \label{line:victim}
    \While{\upshape \code{cohort[other].qIsLocked()} \textbf{or} \code{victim} $ = $ \code{id}}{
    \label{line:petersons:wait}
    \textbf{wait}} 
}
\end{algorithm}

The steps of ALock's \code{Lock(rdma\_ptr<ALock>)} operation (Alg.~\ref{alg:lock} Line~\ref{line:lock}) for a remote access are as follows. First, the thread announces its interest in obtaining the cohort lock by swapping its descriptor onto the remote tail ($tail\_r$) of the lock (Line ~\ref{line:swap}). 
Assuming that this is the first time the ALock is ever acquired, the rCAS on Line~\ref{line:swap} will succeed on its first attempt. 
Otherwise, if the queue was not empty, the rCAS uses the learned value, $prev$, to retry the rCAS and swap the descriptor onto the tail of the queue.
The MCS queue designed for the remote cohort now only contains this thread's descriptor.
As mentioned earlier, a thread requesting a local lock performs the same steps to acquire the ALock but uses shared-memory APIs to interact with the local tail ($tail\_l$), compete to become the leader of the local cohort, and participate in Peterson's algorithm to synchronize with the leader of the remote cohort.

\textbf{Unlock Procedure}.
Unlocking the ALock is done by invoking the \code{Unlock(rdma\_ptr<ALock>)} operation. The thread first attempts to remove its descriptor from the appropriate (i.e., local or remote) tail of the MCS queue. 
Under the assumption that no other concurrent lock request occurres, the modified tail now points to \code{NULL}, which means the thread is leaving the critical section defined by Peterson's algorithm and thus indicates the ALock is successfully unlocked.
On the other hand, if another thread has issued a lock request to the ALock in the meantime, the tail will no longer be equal to the thread's descriptor. 
Following the original MCS algorithm, releasing the lock when a successor is present requires notifying the successor of the lock release by writing to a special state field within the successor's descriptor. 
As mentioned earlier, we refer to notifying the successor of the lock release as \textit{passing} the lock.
This process is further described in Section~\ref{sec:mcs}.

\textbf{Adding Fairness}.
As described so far, the above algorithm is unfair because the lock may be passed indefinitely among threads of the same cohort.
To address fairness, we introduce a budget policy that uses a counter to decide when a cohort must release the ALock. 
To enforce the budget policy, we extend the classic Peterson's algorithm to support a reacquire operation.
With this operation, a thread releases the lock by first setting its cohort as the victim and then immediately attempts to reacquire the lock. 
In the following sections, we refer to the counter variable \code{budget} as an indicator of whether a lock should be released or passed to the waiting successor.
Specifically, if the budget is 0, the lock must be released regardless of a present waiting request. 
Since the lock is released after a bounded number of cohort lock acquisitions, and the lock is itself fair (i.e., a waiting thread cannot be overtaken), our approach is fair~\cite{dice2012lock}.

\begin{figure*}[ht]
    \centering
    \includegraphics[scale=0.45]{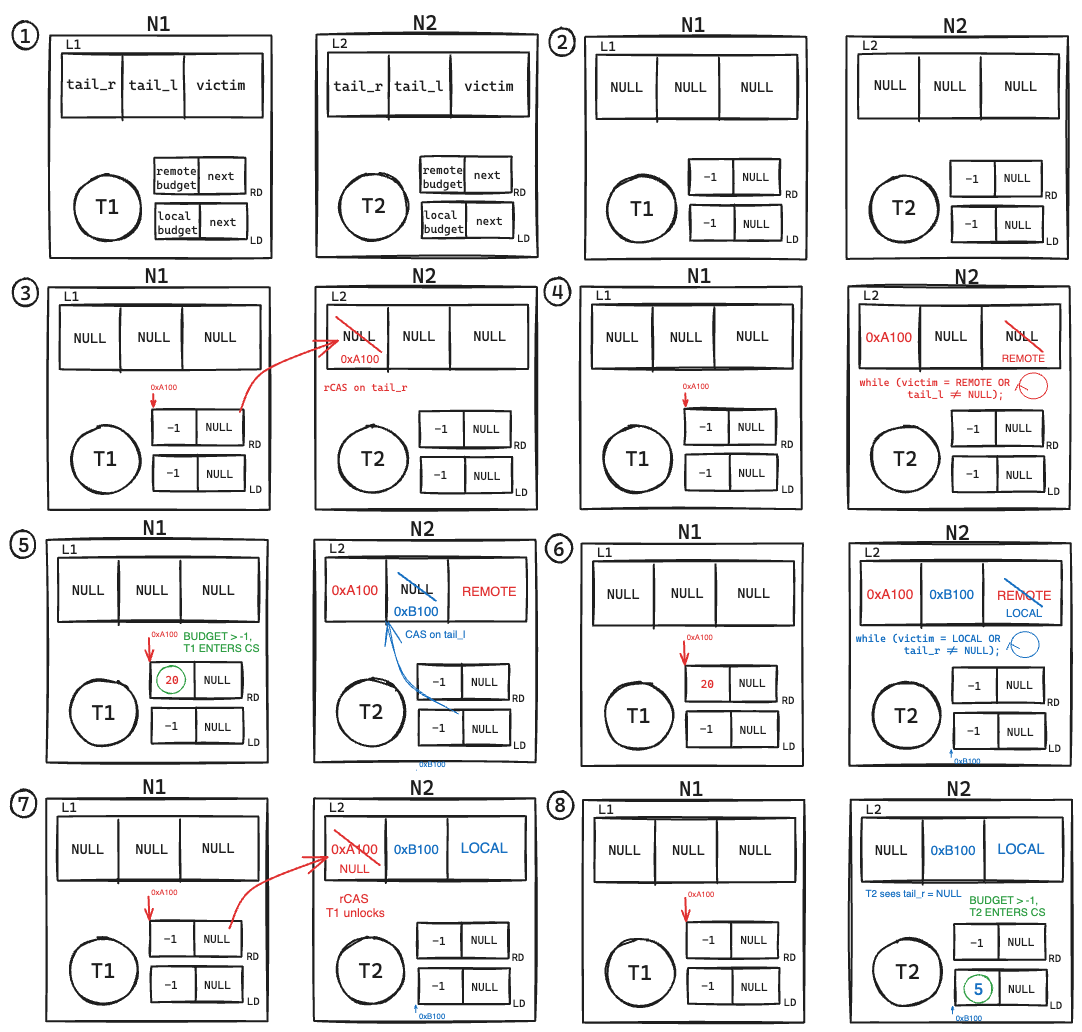}
    \caption{Execution of the ALock algorithm in a system with two nodes, one lock per node, and one thread per node.}
    \label{fig:example}
\end{figure*}

\textbf{Example}.
For the sake of clarity, we define an example to show the procedure that two competing threads of opposite cohorts follow in order to synchronize on the ALock.
We model a cluster with two nodes, $n_1$ and $n_2$. 
One lock resides on each node, $l_1$ on $n_1$ and $l_2$ on $n_2$. 
We also have one thread on each node, $t_1$ and $t_2$. 
For $t_1$, all access requests made on $l_1$ will be considered local, and all access requests made on $l_2$ will be considered remote. 
Likewise, for $t_2$, access requests made on $l_2$ will be considered local, and access requests made on $l_1$ will be considered remote. 
In Figure~\ref{fig:example}, we show the steps taken by $t_1$ to lock $l_2$ followed by $t_2$ attempting to lock $l_2$ concurrently. 

In the first frame \textcircled{1}, we show the 2-node cluster as described with one lock and one thread on each node. 
As shown in frame \textcircled{2}, each thread has a remote descriptor (RD) and local descriptor (LD) that are both initialized with -1 and \code{NULL} values for the budget and next pointer, respectively. 
The first step taken by $t_1$ to lock $l_2$ is to perform a rCAS on the tail of its cohort's MCS queue, which in this case is the remote cohort. 
After successfully performing a rCAS on the remote tail (Frame \textcircled{3}), the thread engages in Peterson's algorithm as the queue was empty when the thread enqueued its descriptor. 
As shown in Frame \textcircled{4}, $t_1$ sets the victim variable to be its cohort (REMOTE), and then enters the while loop, which checks that either the victim is no longer set to REMOTE, or that the local tail's cohort is unlocked. 
In this example, the local tail is currently \code{NULL} (unlocked), so $T_1$ immediately acquires Peterson's lock. 
After seeing that its budget is now non-negative, $t_1$ learns that it has acquired the lock and enters its critical section (Frame \textcircled{5}).

While $t_1$ is in its critical section, $t_2$ also attempts to lock $l_2$. 
In Frame \textcircled{5}, we see $t_2$ locks its cohort by swapping its local descriptors address into the local tail. 
Again, since the queue is empty when the thread adds itself, it must engage in our Peterson's algorithm.
Frame \textcircled{6} shows $t_2$ first sets the victim to be LOCAL, then spins while the victim is still LOCAL or the remote tail is locked (i.e., not null). 
In Frame \textcircled{7}, we see that $t_1$ unlocks $l_2$ by performing an rCAS on the remote tail to set it back to \code{NULL}, removing its descriptor from the queue. 
At this point, $t_2$ will exit the while loop since the remote tail is now unlocked, and upon seeing its now non-negative budget (Frame \textcircled{8}), $t_2$ has acquired the lock and is able to enter its critical section. 

In summary, our technique enables local and remote threads to synchronize via two independent locks integrated into the hierarchical ALock algorithm.
By using a budgeted MCS queue lock as the embedded cohort lock, we can provide fairness.
Lastly, our approach is RDMA-aware since threads performing local accesses avoid RDMA loopback, and threads performing remote accesses limit remote spinning, completely avoiding it in the case competing for Peterson's lock is not required.

\subsection{ALock's MCS lock} \label{sec:mcs}
In this section, we describe the modifications we made to the original MCS queue algorithm for remote accesses.
It should be noted that the local version of the algorithm can be obtained by directly replacing each remote operation with a local one. 
This algorithm maintains a global variable, \code{$tail\_r$}, which is a remote reference to the corresponding slot in the \code{cohort} array of Algorithm~\ref{algo:cohorted} that acts as the tail of the lock queue.

In \code{qLock()}, a thread atomically swaps a new descriptor into \code{$tail\_r$} and then waits until the thread is at the head of the queue, finally returning whether the queue was empty at its onset.
The value swapped into the tail of the queue contains an address of the remotely accessible descriptor (\code{desc}), which is the requesting thread's local metadata.
If \code{$tail\_r$} was not previously set (i.e., was \code{NULL}), the call to \code{qlock()} on Line~\ref{line:qlock} of Algorithm~\ref{alg:lock} returns \code{false} since the cohort lock was not passed to the calling thread.
Otherwise, if the tail was not \code{NULL}, the thread performs an \code{rWrite} operation to the current tail's \code{next} pointer containing the address of the thread's \code{RemoteDescriptor}. 
The locking thread then spins locally on its budget field shown in Line~\ref{line:wait} of Algorithm~\ref{algo:rmcs} while waiting for its predecessor to pass the lock. 
Once the budget field is set to a non-negative value via a \code{rWrite} operation by the predecessor, the lock is considered to be passed to the successor. 

After acquiring the lock and performing its critical section, a thread attempts to release the lock following the conventional MCS queue algorithm, which tries to \code{rCAS} the tail of the queue back to a \code{NULL} value.
Recall that if the \code{rCAS} operation in \code{qUnlock()} is successful, then it also releases the Peterson's lock, since the corresponding \code{cohort} is now unset.
Otherwise, the thread passes the lock to the next waiting thread by performing a \code{rWrite} to the location returned by the attempted \code{rCAS}.
At worst, each \code{Unlock()} operation requires an \code{rCAS} operation followed by an \code{rWrite}.

To support our fairness policy, we alter the original MCS queue algorithm to support a budget, similar to the technique used by Dice et al.~\cite{dice2012lock}.
A lock is passed during the \code{Unlock()} operation by setting the budget of a waiting thread's descriptor to a non-negative integer that represents the number of remaining lock acquisitions.
When the budget reaches zero, a requesting thread must call \code{pReacquire()} on the ALock.
If there is a waiting thread of the opposite cohort, it will be allowed to proceed when the thread sets itself to be the victim (Algorithm \ref{algo:cohorted} Line \ref{line:victim}).
Otherwise, the calling thread reacquires the ALock and resets the budget.

As mentioned earlier, the choice of the MCS algorithm is in part due to its inherent reduction of network contention.
When the queue is empty, a lone thread requires a single \code{rCAS} to acquire the MCS lock, followed by a single \code{rRead} to participate in Peterson's algorithm.
Otherwise, if the queue is not empty, the calling thread spins on its local descriptor, avoiding remote spinning and thus reducing network traffic.

\subsection{ALock's Peterson's Algorithm} \label{sec:petersons}
Our modified Peterson's lock algorithm (Algorithm~\ref{algo:cohorted}) has two global variables: \code{cohort}, which is a two-element array of cohort locks, and \code{victim}, which determines the cohort that yields execution. 
We replace the flag variables in the classic Peterson's algorithm with this array of cohort locks. 
Threads engage in Peterson's algorithm after becoming the leader of their cohort's MCS Queue if they are \textit{1)} the only thread currently requesting the lock; or \textit{2)} passed a budget of 0, requiring the cohort to release the ALock. 

In \code{pReacquire()}, a thread first yields the lock to the waiting cohort by setting the \code{victim} to be its own cohort. 
Recall that a thread's cohort is defined for each operation as remote or local based on whether the access for the lock is remote or local. 
The wait condition (Line~\ref{line:petersons:wait}) in our modified algorithm is similar to the original Peterson's algorithm. 
Checking if the competing thread has finished executing is done by checking whether the other MCS queue cohort has been unlocked. 
As mentioned in Section~\ref{sec:model}, the memory semantics of RDMA is \textit{not} sequentially consistent. 
This requires adding atomic thread fencing instructions after locking and before unlocking and using atomic load and stores for shared memory operations.
Crucially, any interleaving of instructions in concurrent calls to \code{Lock(rdma\_ptr<ALock>)} will only allow a single thread access to the critical section, assuming that sequential consistency is enforced.

\section{Evaluation}
In this section, we present the performance results of our ALock. 
As competitors, we ported two well-known locking algorithms, spinlock, and MCS lock, to RDMA. 
The former simply repeats RDMA \code{rCAS} until it succeeds. 
For the latter, we implement an RDMA-aware queue and integrate it into the original MCS lock algorithm. 
Both these implementations use RDMA for all their operations, regardless of locality. 
In other words, while ALock only performs RDMA operations on remote memory, the competitors use the local RDMA loopback card to perform RDMA operations on local memory.

All our code is written in C++ and compiled with \code{-std=c++20} at optimization level -O3. All competitors are optimized for RDMA operations. 
Each metadata is padded to a size of 64 bytes in order to prevent false cache-line sharing (pictured in Figure~\ref{fig:alock} for ALock), and RDMA pointers remain small at 8 bytes to be friendly to RDMA atomic operations.
We use \code{rdma\_ptr<T>} to represent an RDMA pointer to an object of type T, which is located in RDMA-accessible memory. 
The first 4 bits of the pointer embed the node ID where the memory resides, followed by 60 bits to represent the memory address on that node.
For our experiments, we used 20 machines from the CloudLab~\cite{riccieide2014cloudlab} testbed running on Ubuntu 22.04. 
Each machine is equipped with an Intel Zeon E5-2450 processor with 8/16 cores/threads and one dual-port Mellanox ConnectX-3 RNIC. 

\begin{figure}[h]
    \centering
    \begin{tikzpicture}
        [
            box/.style={rectangle,draw=black, ultra thick, minimum size=1cm},
        ]
    
    \foreach \y [count=\x] in {$tail_r$, $tail_l$, victim}
    {\node[box] at (\x-1,0){\y};}
    \draw[decorate,decoration={brace,mirror},thick] (-.5,-.7) -- node[below]{64B} (2.5,-.7);
    \draw[->,very thick] (-0.5, 1.2) --  node[above,yshift=2mm]{0x0} (-0.5, .7);
    \draw[->,very thick] (0.5,1.2) --  node[above,yshift=2mm]{0x10} (0.5,.7);
    \draw[->,very thick] (1.5,1.2) --  node[above,yshift=2mm]{0x20} (1.5,.7);
    \draw[->,very thick] (2.5,1.2) --  node[above,yshift=2mm]{0x40} (2.5,.7);
    \end{tikzpicture}
    \caption{64B-aligned ALock containing 8B pointers to the remote and local cohort tails, and an integer victim field to indicate the current victim cohort. Values are padded to the address alignments shown.}
    \label{fig:alock}
\end{figure}
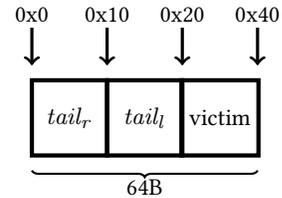 

To test our competitors, we implement a distributed lock table in which locks are partitioned equally across nodes.
We measure the throughput and latency of operations that encompass both one lock and one unlock operation.
We vary the logical contention by changing the size of the lock table (i.e., the number of locks). 
In our experiments, we use 20 locks for a high-contention scenario, 100 for medium contention, and 1000 for low contention. 
We also experiment with different workload levels by varying the number of application threads per node, as well as the number of nodes.
Lastly, we also vary the percentage of operations performed by application threads on local locks. For example, when we refer to 95\% locality, it means 95\% of the operation a thread performs during the experiment target locks that are stored in the same node where the threat executes (i.e., local accesses as defined in Section~\ref{sec:model}).

We conduct our study by first analyzing the impact of the budget on the performance of ALock. 
The results of these experiments inform our decision to choose a remote and local budget for our remaining experiments.
We then continue by measuring throughput and latency under different conditions.

\subsection{Choosing a Budget} \label{sec:budget}
As described in Section~\ref{sec:alock}, the ALock algorithm provides fairness through the use of a budget.
However, the budget has a double effect: preventing starvation while also minimizing the cost of the lock reacquire operation. 
There is an \textit{asymmetric} cost associated with the reacquire operation for remote and local threads because a local thread needs just shared-memory operations to participate in the reacquire, while a remote thread needs RDMA operations.
For this reason, we have two separately defined budget configurations: one local budget and one remote budget(see Algorithm~\ref{alg:structs}).

Our intuition is that keeping the local budget low minimizes the time that a thread requesting a remote lock may spin in the high-contention workload scenario. 
On the other hand, threads that are requesting a remote lock and waiting for it to be released benefit from a higher remote budget as it reduces the number of times the reacquire operation is invoked.

\begin{figure}[ht]
    \centering
\includegraphics[width=0.75\linewidth]{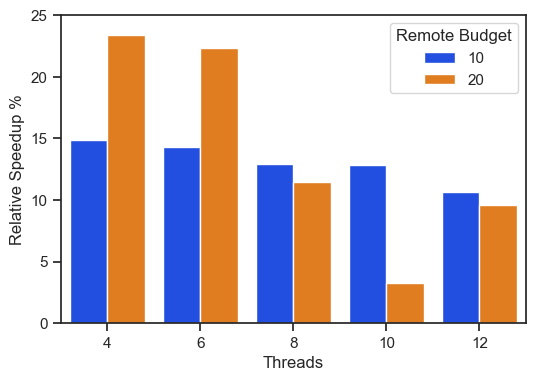}
    \caption{Relative speedup compared to the baseline remote budget of 5.}
    \label{fig:budget}
\end{figure}

\begin{figure*}[ht]
    \centering
    \begin{subfigure}{.25\textwidth}
        \centering
        \includegraphics[width=\textwidth]{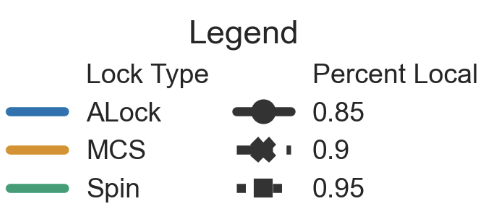}
    \end{subfigure}
    \begin{subfigure}{.19\textwidth}
        \hspace*{4.2cm}
        \includegraphics[width=\textwidth]{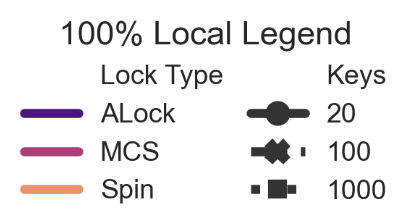}
    \end{subfigure}\\
    \begin{subfigure}{\textwidth}
        \centering
        \includegraphics[scale=0.65]{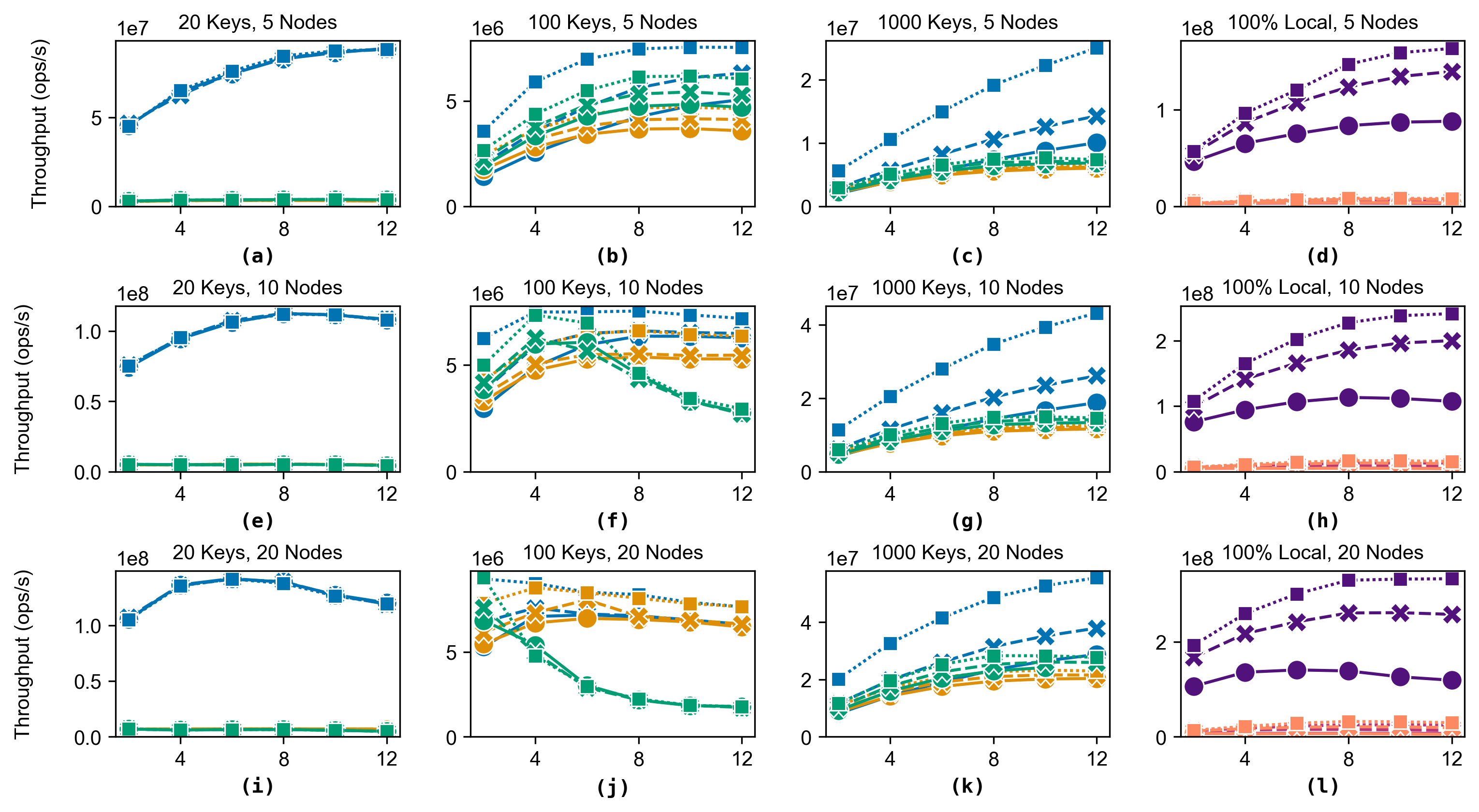}
    \end{subfigure}
    \caption{Throughput for a variety of workloads on 5, 10, and 20 nodes. The X-axis is the number of threads per node.}
    \label{fig:tputplots}
\end{figure*}

We performed a series of experiments with various workloads and cluster configurations to confirm our intuition and inform our choice for both the local and remote budgets. 
We found that the most impactful factor for the budget is the contention as provided by the workload.
Figure~\ref{fig:budget} shows the average performance improvement relative to a baseline configuration having both remote and local budgets set to 5. 
In the plot, we show ALock's performance while increasing the remote budget from 5 to 10 and 20.
Results are averaged under 95\%, 90\%, and 85\% locality workloads on a 20-node cluster with 100 locks (medium contention).

As shown by the plot, limiting the local budget while increasing the remote budget improves total throughput by up to 23\%. We attribute this to the fact that competing in the \code{Reacquire} operation is much more costly for threads requesting remote locks rather than local locks due to the cost of the remote spinning encountered in Peterson's algorithm when contention from the local cohort is present. 
Even without local contention, the cost of two additional remote operations at the minimum needed for the remote cohort to reacquire is much more costly than the cost of two additional local operations needed for the local cohort to reacquire \cite{kalia2016design}.  
These findings informed us to choose a remote budget of 20 and a local budget of 5 for the rest of the experiments.

\begin{figure*}[t]
    \centering
    \begin{subfigure}{.4\textwidth}
        \centering
        \includegraphics[width=\textwidth]{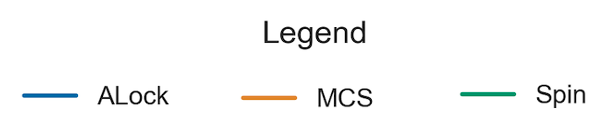}
    \end{subfigure}
    \begin{subfigure}{\textwidth}
        \centering
        \includegraphics[scale=0.75]{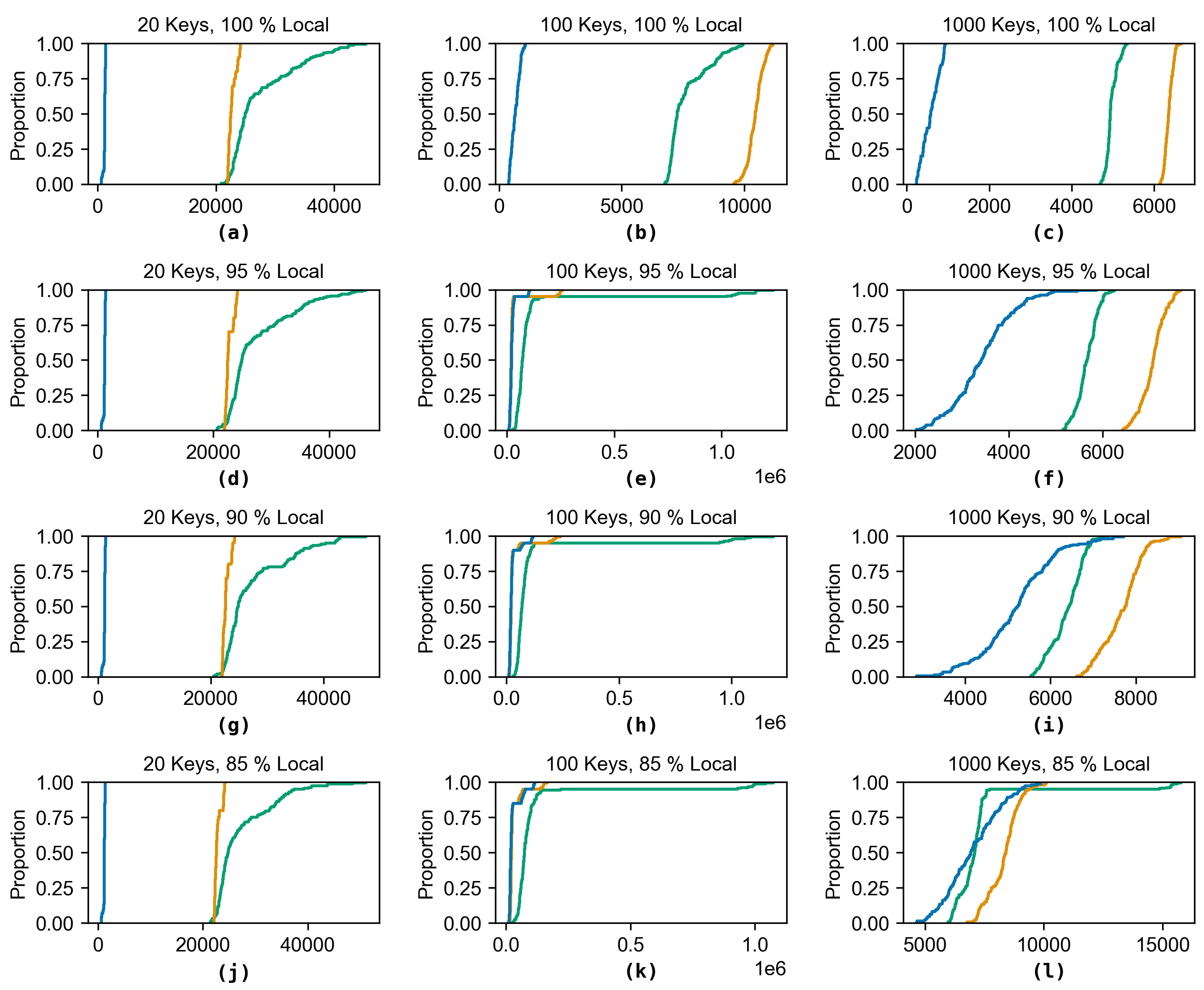}
    \end{subfigure}
    \caption{Latency CDF for a 10 node cluster with 8 threads under different workloads. The X-axis is latency in nanoseconds.}
    \label{fig:latplots}
\end{figure*}

\subsection{Throughput Evaluation}
In Figure~\ref{fig:tputplots}, we report the throughput in terms of operations (i.e., one successful lock and unlock) per second. We vary the size of the system, the workload, and the contention level.

We first discuss the case of 100\% locality. 
Because of the significant gap in performance, we isolated this workload on separate plots (Figures (d), (h), (l)).
As evident by the plots, the ALock significantly outperforms the spinlock and MCS lock at low, medium, and high contention levels. 
Specifically, the ALock performs up to 24x as many operations as the MCS Lock and 22x as many operations as the spinlock, even under high contention with just 20 locks. 
Our ALock is able to take full advantage of the beneficial operation asymmetry here and \textit{only} use shared-memory operations. 
Additionally, there is no contention in Peterson's algorithm due to the absence of the remote cohort.

Moving away from 100\% locality, we focus on the high-contention case (Figures (a), (e), (i)).
The spinlock and MCS lock are both overwhelmed in these cases.
The most extreme contention is in Figure (i).
With up to 240 threads competing for just 20 locks in the 20-node configuration, the workload creates congestion in the RNIC, which quickly degrades performance. 
Conversely, the ALock outperforms the MCS lock by up to 29x and the spinlock by up to 24x. 
Avoiding the performance bugs introduced by loopback traffic allows the ALock to take full advantage of the passing lock mechanisms. 

As the contention decreases, ALock continues to perform well, but competitors' performance increases, resulting in a smaller gap (Figures (b), (f), (j)). 
Changing from 20 to 100 locks, the MCS lock is no longer overwhelmed by the logical contention and, like the ALock, takes advantage of spinning locally and passing the lock as the workload scales.
Interestingly, and somehow counter-intuitive, although the ALock maintains its scalability, it experiences a decrease in throughput as the contention decreases. 
This is due to the fact that passing the lock cannot always be extensively used without enough contention.

The spinlock, however, saturates quickly and suffers as the workload scales.
Since the spinlock uses remote spinning, the card's congestion increases with the number of threads, causing undesirable performance.
We attribute this behavior to the congestion introduced by the loopback mechanism, and QP thrashing.
In fact, with 8 threads and 10 nodes, the RNIC already needs to maintain 1280 QPs.
However, recent work has shown that performance degrades as the number of concurrent connections grows past 450 \cite{wang2021star}.
In comparison, the ALock throughput scales with the number of threads as it avoids loopback and limits QP thrashing, outperforming the simple spinlock approach by more than 4x with 240 total threads in the system (Figure (j)).

Finally, as contention decreases (Figures (c), (g), (k)), the ALock increases its performance gains as the workload's locality increases.
Without logical contention, passing the lock becomes less relevant, but the asymmetry cost of remote and local operations becomes more evident. 
The ALock improves by 40\% when increasing from 85\% to 90\% locality and improves by an additional 75\% when increasing to 95\% locality using five nodes. 
The spinlock and MCS lock are also able to scale their performance with the number of threads but at a slower pace than ALock. However, they do not experience the performance benefits that ALock can from the use of shared-memory operations as the workload locality increases.
In this case, the ALock outperforms the MCS lock by up to 3.8x and the spinlock by up to 3.3x.

\subsection{Latency Evaluation}
In Figure~\ref{fig:latplots}, we report the cumulative distribution of the latency of operations (i.e., one lock and unlock).
In our experiments, we varied the workload, contention level, and size of the system.
We chose only to include results for the 20-node cluster in order to show the latency at scale. That is because there are minimal differences in the plots with 5 and 10 nodes, except for the performance of spinlock, whose scalability can be better assessed with 20 nodes.

In the 100\% local workloads (Figures (a), (b), (c)), the ALock significantly outperforms its competitors. 
Since the ALock is composed of 100\% local operations, the latency is close to that of local memory accesses.
Specifically, in the high-contention case with 20 locks (Figure (a)), the ALock is as much as 17x faster than the MCS lock and 33x faster than the spinlock.
As the number of locks increases, ALock's tail latency also increases since the passing of the lock can no longer be always used as the contention decreases. 
On the contrary, the spinlock experiences a much more drastic tail in the high-contention cases with 20 locks (Figures (a), (d), (g), (j)) as it is overwhelmed by the logical contention, causing congestion in the card.
The MCS lock, similar to the ALock, takes advantage of spinning locally in these high-contention cases to avoid such a large tail.
However, since it still requires many RDMA operations, the MCS lock competitor still experiences a latency of up to 17x longer than the ALock under high contention.
In low contention and 100\% locality (Figure (c)), the ALock latency is still an average of 10x faster than spinlock and 13x faster than MCS lock.

Moving away from 100\% locality and looking at the case of medium contention with 100 locks (Figures (e), (h), (k)), we see that ALock and MCS lock, which share a similar structure, perform very similarly.
In these workloads, the logical contention is enough for both the ALock and MCS lock to take advantage of passing the lock and spinning locally, allowing the MCS lock to avoid saturation. 
On the other hand, the spinlock's long tail latency can be attributed to the congestion caused by remote spinning. 

In the case of low contention (Figures (f), (i), (l)), ALock outperforms MCS by an average of 2.1x in the 95\% local workload and 1.35x in the 85\% local. 
With the absence of logical contention, both ALock and MCS lock are unable to benefit from passing.
However, being unable to pass the lock results in the MCS lock requiring many more remote operations, whereas the ALock can benefit from the locality of the workload. 
This is evident when looking at Figures (i) and (l).
As the locality decreases from 90\% to 85\%, the gap between the ALock and MCS lock's performance shrinks.

\section{Related Work}
Mutual exclusion is a known problem in which access to a shared resource is coordinated among two or more concurrent threads~\cite{lynch1996distributed}.
Specifically, at most one thread may execute its critical section at a time.
A naive solution to multi-thread mutual exclusion is a \emph{filter lock}~\cite{peterson1981myths}, which extends Peterson's lock for multiple threads.
Briefly, threads compete for access to successive levels that each hold back one thread.
The number of levels is equal to one less than the number of threads that might acquire the lock.
Unfortunately, this would require both remote spinning and a number of remote operations proportional to the number of threads that might contend for the lock, even if a thread executes in isolation.
Lamport's Bakery algorithm~\cite{herlihy2021} also demonstrates the same undesirable behavior for remote threads. 

Similar to lock cohorting~\cite{dice2012lock}, a strategy for NUMA-aware synchronization, our approach allows a group of threads to compete amongst themselves before acquiring a global lock, in our case the ALock.
Our technique explicitly couples the Peterson's and MCS locks to achieve behavior tailored to the respective access types in our system, remote and local.
By embedding the cohort locks in Peterson's lock, locking and unlocking the MCS queue simultaneously sets and un-sets the Peterson's flag variable. 
As a result, we avoid an additional remote operation for remote accesses while maintaining the integrity of the ALock.
The application of lock cohorting in a distributed setting is a natural extension of the technique but it requires rethinking the design to optimize for operation asymmetry between local and remote accesses, yielding a lock primitive that is of independent interest.

To the best of our knowledge, our approach is the first mutual exclusion primitive designed for RMDA that provides local-only access for threads requesting local locks while maintaining fairness and avoiding RPCs.
A notable alternative is the technique pioneered by Wei et al.~\cite{wei2020fast}, which allows local accesses to be protected by hardware transactional memory (HTM) while remote accesses acquire a lock using RDMA \code{rCAS}.
This technique only applies to architectures supporting HTM, which is increasingly disabled due to security concerns~\cite{lipp2018meltdown, intel2021performance}.
Due to cache coherent I/O, a local hardware transaction is aborted whenever a remote thread acquires the lock.
Local operations use local accesses in the common case, but a fallback path using RDMA is also required.
Another potential option is to leverage RDMA-accessible memory permissioning, which atomically revokes remote access~\cite{aguilera2019impact,2007infiniband}, to devise a mutual exclusion algorithm.
However, this approach is known to be slow~\cite{aguilera2020microsecond} and is not easily made starvation-free since remote access may be continuously revoked by threads performing local accesses.

Recently, new cache coherent interconnects like CXL have attracted attention and are expected to play a role in implementing disaggregated memory patterns. 
Cache coherence can make it possible to use both RDMA and local atomic operations without the need for an additional synchronization mechanism.
However, to fully take advantage of CXL will require RNIC redesign and may still come with a performance tradeoff for coherency. At the current state of the technology (not yet released), these considerations are mostly speculation~\cite{gouk2022DirectCXL}.

\section{Conclusion}
In this paper, we face the challenge of synchronizing accesses in RDMA-based systems and define operation asymmetry to capture the disparate behavior of how remote and local threads operate on RDMA memory. 
We propose a fair, starvation-free mutual exclusion primitive that enables local and remote accesses to synchronize globally in a manner that is abstracted away from the programmer while optimizing for their individual characteristics.
Inspired by lock cohorting, we embed RDMA-aware MCS locks into a modified version of the well-known Peterson's algorithm to create the ALock.
To the best of our knowledge, our technique is the first mutual exclusion solution that allows synchronizing local and remote accesses while avoiding known RDMA scalability issues (abuse of RDMA loopback and RPCs).

\begin{acks}
This material is based upon work supported by the National Science Foundation under Grant No. CNS-2045976. This research was also funded by a CORE grant from Lehigh University and by a gift grant from the Stellar Dev. Foundation.
\end{acks}

\bibliographystyle{ACM-Reference-Format}
\bibliography{references}

\clearpage

\appendix
\section*{APPENDIX}
\section{TLA+ Specification}
\label{app:tla}

\definecolor{boxshade}{gray}{0.85}

\tlatex
\small
\setboolean{shading}{true}
\@x{}\moduleLeftDash\@xx{ {\MODULE} alock}\moduleRightDash\@xx{}%
\@x{ {\EXTENDS} Integers ,\, Sequences ,\, TLC}%
\@x{ {\CONSTANTS} NumProcesses ,\, InitialBudget}%
\@x{ {\ASSUME} NumProcesses \.{>} 0}%
\@x{ {\ASSUME} InitialBudget \.{>} 0}%
\@x{ NP \.{\defeq} NumProcesses}%
\@x{ B \.{\defeq} InitialBudget}%
\pcalsymbolstrue
\csyntaxfalse
\@x{\@s{8.06} {\p@mmalgorithm} alock}%
\@x{ {\p@variables}}%
\@x{\@s{16.4}}%
\@y{\@s{0}%
 Global
}%
\@xx{}%
\@x{\@s{16.4} victim\@s{0.30} \.{\in} \{ 1 ,\, 2 \} ,\,}%
\@x{\@s{16.4} cohort \.{=} [ x \.{\in} \{ 1 ,\, 2 \} \.{\mapsto} 0 ] ,\,}%
 \@x{\@s{16.4} descriptor \.{=} [ x \.{\in} ProcSet \.{\mapsto} [ budget
 \.{\mapsto} \.{-} 1 ,\, next \.{\mapsto} 0 ] ] ,\,}%
\@x{\@s{16.4}}%
\@y{\@s{0}%
 Process-local
}%
\@xx{}%
\@x{\@s{16.4} passed \.{=} [ x \.{\in} ProcSet \.{\mapsto} {\FALSE} ] ,\,}%
\@vspace{8.0pt}%
\@x{ {\p@define}}%
\@x{\@s{16.4} Us ( pid ) \.{\defeq} ( pid \.{\%} 2 ) \.{+} 1}%
\@vspace{2.0pt}%
\@x{\@s{16.4} Them ( pid ) \.{\defeq} ( ( pid \.{+} 1 ) \.{\%} 2 ) \.{+} 1}%
\@vspace{2.0pt}%
\@x{\@s{16.4} Budget ( pid ) \.{\defeq} descriptor [ pid ] . budget}%
\@x{ {\p@end} {\p@define} {\p@semicolon}}%
\@vspace{8.0pt}%
\@x{ {\p@procedure} AcquireGlobal ( )}%
\@x{ {\p@begin}}%
\@x{\@s{16.4} g1\@s{.5}\textrm{:}\@s{3} victim \.{:=} self {\p@semicolon}}%
\@x{\@s{16.4} gwait\@s{.5}\textrm{:}\@s{3} {\p@while} {\TRUE} {\p@do}}%
 \@x{\@s{50.47} g2\@s{.5}\textrm{:}\@s{3} {\p@if} cohort [ Them ( self ) ]
 \.{=} 0 {\p@then}}%
\@x{\@s{81.87} {\p@goto} g4 {\p@semicolon}}%
\@x{\@s{67.23} {\p@end} {\p@if} {\p@semicolon}}%
 \@x{\@s{50.47} g3\@s{.5}\textrm{:}\@s{3} {\p@if} victim \.{\neq} self
 {\p@then}}%
\@x{\@s{81.87} {\p@goto} g4 {\p@semicolon}}%
\@x{\@s{67.23} {\p@end} {\p@if} {\p@semicolon}}%
\@x{\@s{33.16} {\p@end} {\p@while} {\p@semicolon}}%
\@x{\@s{16.4} g4\@s{.5}\textrm{:}\@s{3} {\p@return}}%
\@x{ {\p@end} {\p@procedure} {\p@semicolon}}%
\@vspace{8.0pt}%
\@x{ {\p@procedure} AcquireCohort ( )}%
\@x{ {\p@variables} pred}%
\@x{ {\p@begin}}%
 \@x{\@s{16.4} c1\@s{.5}\textrm{:}\@s{3} descriptor [ self ] \.{:=} [ budget
 \.{\mapsto} \.{-} 1 ,\, next \.{\mapsto} 0 ] {\p@semicolon}}%
 \@x{\@s{16.4} swap\@s{.5}\textrm{:}\@s{3} pred \.{:=} cohort [ Us ( self ) ]
 {\p@semicolon} cohort [ Us ( self ) ] \.{:=} self {\p@semicolon}}%
 \@x{\@s{16.4} cwait\@s{.5}\textrm{:}\@s{3}\@s{1.99} {\p@if} {\lnot} ( pred
 \.{=} 0 ) {\p@then}}%
 \@x{\@s{32.8} c2\@s{.5}\textrm{:}\@s{3}\@s{3.22} descriptor [ pred ] . next
 \.{:=} self {\p@semicolon}}%
 \@x{\@s{32.8} c3\@s{.5}\textrm{:}\@s{3}\@s{3.22} {\p@await} Budget ( self )
 \.{\geq} 0 {\p@semicolon}}%
 \@x{\@s{32.8} c4\@s{.5}\textrm{:}\@s{3}\@s{3.22} {\p@if} Budget ( self )
 \.{=} 0 {\p@then}}%
 \@x{\@s{67.10} c5\@s{.5}\textrm{:}\@s{3} {\p@call} AcquireGlobal ( )
 {\p@semicolon}}%
 \@x{\@s{67.10} c6\@s{.5}\textrm{:}\@s{3} descriptor [ self ] . budget \.{:=}
 B {\p@semicolon}}%
\@x{\@s{52.46} {\p@end} {\p@if} {\p@semicolon}}%
 \@x{\@s{32.8} c7\@s{.5}\textrm{:}\@s{3}\@s{3.22} passed [ self ] \.{:=}
 {\TRUE} {\p@semicolon}}%
\@x{\@s{16.4} {\p@else}}%
 \@x{\@s{32.8} c8\@s{.5}\textrm{:}\@s{3} descriptor [ self ] . budget \.{:=} B
 {\p@semicolon}}%
 \@x{\@s{32.8} c9\@s{.5}\textrm{:}\@s{3} passed [ self ] \.{:=} {\FALSE}
 {\p@semicolon}}%
\@x{\@s{16.4} {\p@end} {\p@if} {\p@semicolon}}%
\@x{\@s{16.4} c10\@s{.5}\textrm{:}\@s{3}\@s{4.53} {\p@return} {\p@semicolon}}%
\@x{ {\p@end} {\p@procedure} {\p@semicolon}}%
\@vspace{8.0pt}%
\@x{ {\p@procedure} ReleaseCohort ( )}%
\@x{ {\p@variables} size ,\, next}%
\@x{ {\p@begin}}%
 \@x{\@s{16.4} cas\@s{.5}\textrm{:}\@s{3} {\p@if} cohort [ Us ( self ) ] \.{=}
 self {\p@then}}%
\@x{\@s{32.8} cohort [ Us ( self ) ] \.{:=} 0 {\p@semicolon}}%
\@x{\@s{16.4} {\p@else}}%
 \@x{\@s{32.8} r1\@s{.5}\textrm{:}\@s{3} {\p@await} {\lnot} ( descriptor [
 self ] . next \.{=} 0 ) {\p@semicolon}}%
 \@x{\@s{32.8} r2\@s{.5}\textrm{:}\@s{3} descriptor [ descriptor [ self ] .
 next ] . budget \.{:=} Budget ( self ) \.{-} 1 {\p@semicolon}}%
\@x{\@s{16.4} {\p@end} {\p@if} {\p@semicolon}}%
\@x{\@s{16.4} r3\@s{.5}\textrm{:}\@s{3}\@s{5.31} {\p@return} {\p@semicolon}}%
\@x{ {\p@end} {\p@procedure} {\p@semicolon}}%
\@vspace{8.0pt}%
\@x{ {\p@fair} {\p@process} p \.{\in} 1 \.{\dotdot} NP}%
\@x{ {\p@begin}}%
\@x{\@s{16.4} p1\@s{.5}\textrm{:}\@s{3} {\p@while} {\TRUE} {\p@do}}%
\@x{\@s{33.42}}%
\@y{\@s{0}%
 Non-critical section
}%
\@xx{}%
\@x{\@s{33.42} ncs\@s{.5}\textrm{:-}\@s{3} {\p@skip} {\p@semicolon}}%
\@vspace{8.0pt}%
\@x{\@s{33.42}}%
\@y{\@s{0}%
 Acquire the cohort lock
}%
\@xx{}%
 \@x{\@s{33.42} enter\@s{.5}\textrm{:}\@s{3} {\p@call} AcquireCohort ( )
 {\p@semicolon}}%
\@vspace{8.0pt}%
\@x{\@s{33.42}}%
\@y{\@s{0}%
 Acquire the global lock, maybe
}%
\@xx{}%
 \@x{\@s{33.42} p2\@s{.5}\textrm{:}\@s{3} {\p@if} {\lnot} passed [ self ]
 {\p@then}}%
\@x{\@s{67.64} {\p@call} AcquireGlobal ( ) {\p@semicolon}}%
\@x{\@s{50.44} {\p@end} {\p@if} {\p@semicolon}}%
\@vspace{8.0pt}%
\@x{\@s{33.42}}%
\@y{\@s{0}%
 Critical section
}%
\@xx{}%
\@x{\@s{33.42} cs\@s{.5}\textrm{:}\@s{3} {\p@skip} {\p@semicolon}}%
\@vspace{8.0pt}%
\@x{\@s{33.42}}%
\@y{\@s{0}%
 Release the cohort lock
}%
\@xx{}%
 \@x{\@s{33.42} exit\@s{.5}\textrm{:}\@s{3} {\p@call} ReleaseCohort ( )
 {\p@semicolon}}%
\@x{\@s{16.4} {\p@end} {\p@while} {\p@semicolon}}%
\@x{ {\p@end} {\p@process} {\p@semicolon}}%
\@x{ {\p@end} {\p@algorithm}}%
\@y{%
 ;
}%
\@xx{}%
\pcalshadingfalse \pcalsymbolsfalse
\@vspace{8.0pt}%
\@x{}%
\@y{\@s{0}%
 Safety
}%
\@xx{}%

 \@x{ MutualExclusion \.{\defeq} ( \A\, i ,\, k \.{\in} ProcSet \.{:} ( i
 \.{\neq} k ) }%
  \@x{ \.{\implies} {\lnot} ( pc [ i ] \.{=}\@w{cs} \.{\land} pc [ k ]
 \.{=}\@w{cs} ) )}%
\@vspace{8.0pt}%
\@x{}%
\@y{\@s{0}%
 Liveness
}%
\@xx{}%
 \@x{ ExecsCriticalSectionInfinitelyOften \.{\defeq} \A\, i \.{\in} ProcSet
 \.{:} {\Box} {\Diamond} ( pc [ i ] \.{=}\@w{cs} )}%
 \@x{ StarvationFree \.{\defeq} \A\, i \.{\in} ProcSet \.{:} ( pc [ i
 ]\@s{9.22} \.{=}\@w{enter} ) \.{\leadsto} ( pc [ i ] \.{=}\@w{cs} )}%
 \@x{ DeadAndLivelockFree \.{\defeq} ( \E\, i \.{\in} ProcSet \.{:} pc [ i ] \.{=}\@w{enter} )}%
 \@x{ \.{\leadsto} ( \E\, i \.{\in} ProcSet \.{:} pc [ i ] \.{=}\@w{cs} )}%
\@vspace{8.0pt}%
\@x{}%
\@y{\@s{0}%
 Fairness
}%
\@xx{}%
 \@x{ CohortFairness \.{\defeq} \A\, i ,\, j \.{\in} ProcSet \.{:} ( pc [ i ]
 \.{=}\@w{cwait}\@s{0.55} \.{\land} pc [ j ] \.{=}\@w{enter} )}%
 \@x{\.{\implies} (pc [ i ] \.{=}\@w{cs} \.{\leadsto} pc [ j ] \.{=}\@w{cs} )}%
 \@x{ GlobalFairness\@s{2.86} \.{\defeq} \A\, i ,\, j \.{\in} ProcSet \.{:} (
 pc [ i ] \.{=}\@w{gwait} \.{\land} pc [ j ] \.{=}\@w{enter} ) }%
 \@x{\.{\implies} (pc [ i ] \.{=}\@w{cs} \.{\leadsto} pc [ j ] \.{=}\@w{cs} )}%
\@x{}\bottombar\@xx{}%
\endtla

\end{document}